# Nanoscale axial position and orientation measurement of hexagonal boron nitride quantum emitters using a tunable nanophotonic environment

Pankaj K. Jha[1, †], Hamidreza Akbari[1, †], Yonghwi Kim[1, †], Souvik Biswas[1], Harry A. Atwater[1, *]

*[1]Thomas J. Watson Laboratory of Applied Physics and Materials Science, California Institute of Technology, Pasadena, CA 91125, USA.*

*[2]Resnick Sustainability Institute, California Institute of Technology, Pasadena, CA 91125, USA.*

*[3]Joint Center for Artificial Photosynthesis, California Institute of Technology, Pasadena, CA 91125, USA.*

*[†]These authors contributed equally to this work.*

*[*]Corresponding author: Harry A. Atwater (haa@caltech.edu)*

(Dated: February 23, 2021)

**Color centers in hexagonal boron nitride ($h$BN) have emerged as promising candidates for single-photon emitters (SPEs) due to their bright emission characteristics at room temperature. In contrast to mono- and few-layered $h$BN, color centers in multi-layered flakes show superior emission characteristics such as higher saturation counts and spectral stability. Here, we report a method for determining both the axial position and three-dimensional dipole orientation of SPEs in thick $h$BN flakes by tuning the photonic local density of states using vanadium dioxide ($VO_2$), a phase change material. Emitters under study exhibit a strong surface-normal dipole orientation, providing some insight on the atomic structure of $h$BN SPEs, deeply embedded in thick crystals. We have optimized a hot pickup technique to reproducibly transfer flakes of $h$BN from $VO_2$ onto $SiO_2$/Si substrate and relocated the same emitters. Our approach serves as a practical method to systematically characterize SPEs in $h$BN prior to integration in quantum photonics systems.**

**Introduction**

Over the past few decades, point defects *(1)* that introduce electronic states with optical transitions, also known as color centers, have garnered great interest for quantum photonics applications, such as quantum computation and quantum information *(2,3)*, quantum cryptography *(4)*, and quantum sensing *(5)*. Wide-bandgap materials, such as diamond *(6)*, silicon carbide *(7)*, gallium nitride *(8)*, and zinc oxide *(9)* offer promising platforms for hosting quantum emitters with emission in the visible to near-infrared spectrum. However, these materials suffer from one or more intrinsic challenges such as a requirement for cryogenic temperatures, decoherence of emitted photons, optical coupling losses, and challenges associated with chip-based photonic integration. These problems have driven researchers to seek new candidate materials with fewer disadvantages *(10,11)*.

Recent discoveries of quantum light emission from two-dimensional van der Waals (vdW) layered materials *(12-17)* have introduced promising candidates for single photon emitters (SPEs). In contrast to bulk materials, vdW materials offer easier integration with photonic structures and minimal loss due to refractive index mismatch *(18,19)*. Among several candidate vdW host materials, hexagonal boron nitride (*h*BN) has received particular attention due to its ability to offer a bright source of quantum light at room temperature. Remarkably, quantum emitters in *h*BN have shown high (> 80%) Debye-Waller factor *(17)*, a brightness comparable to the brightest SPEs *(10, 19)*, polarized emission *(17, 20)*, giant stark shift *(21-23)*, magnetic-field dependent quantum emission *(24,25)*, correlated cathodoluminescence and photoluminescence emission *(26)*, and near transform-limited optical linewidth *(27)*, all reported at room temperature. To this date, the atomic structure of *h*BN quantum emitters is not clear; the most common approach to deduce the atomic structure of these emitters has been to compare the energy of the zero-phonon line (ZPL) and phonon-assisted emission to the first-principles calculations *(28)*. However, this approach cannot narrow down the pool of possible candidates. Accurate information of 3D orientation of emitting dipole would provide invaluable insight into the underlying symmetry properties of the defect center, which can complement the aforementioned approach and help in identifying the atomic origin of emitters.

Quantum emitters in multi-layered flakes, in contrast to mono- and few-layered flakes, show superior emission characteristics such as higher saturation counts and spectral stability due to reduced environmental screening effects *(29)*. Efficient coupling of these quantum emitters with nanophotonic structures would require precise information about their axial position and 3D dipole orientation. Determining, both, the 3D dipole orientation and axial position of a quantum emitter in any multi-layered $h$BN poses as a coupled problem because the polarization characteristics of detected photons is strongly influence by either dipole orientation or axial position.

In this Letter, we demonstrate nanometer-scale axial location of $h$BN quantum emitters in a multi-layered flake by leveraging highly sensitive, distance-dependent modulation of the spontaneous emission lifetime of these quantum emitters when in close proximity to a tunable phase-change material, vanadium dioxide ($VO_2$). Specifically, we modify the local density of optical states (LDOS) by inducing an insulator-to-metal transition in $VO_2$ which in turn modulates the emission rate of quantum emitters near the $h$BN/$VO_2$ interface. This method, taken together with emission polarimetry to determine the three-dimensional (3D) orientation of the quantum emitter, give comprehensive information about emitters axial position and orientation, which could be used to distinguish possible candidates based on atomic structure of the emitter. Furthermore, we have optimized a polymer-assisted hot pickup and transfer process to reliably transfer $h$BN flakes from $VO_2$ substrate onto any arbitrary substrate (here, we use $SiO_2$/Si substrate as target), following which, the emitters under investigation were relocated. Measurement of $h$BN emitter location and 3D orientation with nanometer-scale resolution in multi-layered flakes together with advances in precise transfer and stacking of 2D materials (*30*) and metal contacts (*31*) offer opportunities for both fundamental physics advances (*32*) and quantum photonic technologies (*33,34*).

The experimental configuration is shown in Figure 1a. A quantum emitter is located at a distance *d*, within the thickness of an $h$BN flake, from the surface of a substrate that consists of a thin layer of $VO_2$ on sapphire. Photoluminescence excitation and detection were performed with optical pumping of quantum emitters by laser excitation from the top. The excited quantum emitter emission decay rate depends on its interaction with the optical environment (*35*). By optical environment we mean the substrate beneath and the air above the $h$BN flake. We model this interaction by treating the quantum emitter as an oscillating point dipole source oriented along the direction $(\theta, \varphi)$. For an emitter in an unbounded, homogeneous, and lossless medium with a refractive index *n*, the spontaneous decay rate is enhanced by a factor *n* compared to the free space. This result also holds true for bounded geometry as long as the emitter is at a distance $d \gg \lambda$ from any interface. When $d \ll \lambda$ the decay rate strongly depends on *d*, the dipole orientation $(\theta, \varphi)$, and the refractive index contrast across the interface (*36-38*). In this work, we manipulated the optical environment of a quantum emitter located in the vicinity of the $h$BN/VO$_2$ interface using VO$_2$ whose complex refractive index exhibits a sharp change when VO$_2$ is thermally switched from the insulating to metallic state, which occurs at near room temperature $T_c \sim 340$ K (*39*).

Figure 1b shows the calculated relative decay rate $\beta = \gamma_{Insulating}/\gamma_{Metallic}$ of an emitter as a function of distance $d$ when the emitter is oriented perpendicular ($\theta = 0°$) and parallel ($\theta = 90°$) to the $h$BN/VO$_2$ interface. Here, $\gamma_{Insulating}$ and $\gamma_{Metallic}$ is the total (radiative and non-radiative) decay rate of the emitter when VO$_2$ is in insulating (30 ℃) and metallic (100 ℃) state respectively. In these simulations, we considered a flake thickness of 310 nm and an emission wavelength of 600 nm, corresponding to one of the quantum emitters in our experiment, shown in Figs. 4a and 4b. The refractive indices of the upper medium, $h$BN, and that of sapphire were set to 1, 1.82 (*40*), and 1.77 respectively. The complex refractive index of VO$_2$ at 600 nm was extracted from our ellipsometric data and was set to 3.05 + 0.42i and 2.57 + 0.64i for VO$_2$ in insulating and metallic state respectively (Supplementary Section S1). The thickness of the VO$_2$ layer was 40 nm. In general, the photoluminescence quantum yield (PLQY) of $h$BN quantum emitters varies in the range 0.6-1.0 (*10,19*) and a recent experiment has shown average PLQY in the range 0.6-0.8 for quantum emitters with zero-phonon line (ZPL) around 600 nm (*41*). The shaded area in Figure 1b, corresponds to this PLQY range. The dashed line within the shaded regions corresponds to a PLQY = 0.79 estimated from our experimental data (Supplementary Section S4). As can be seen, the relative modulation of decay rates for both orientations is clearly evident within first ~50 nm that quickly fades away at distances ~100 nm and above. We use this highly sensitive, distance-dependent decay rate of quantum emitters in the vicinity of the $h$BN/VO$_2$ interface to localize their position along the axial direction.

Figure 2(a) shows the optical microscope image of a thin $h$BN flake on a VO$_2$/Sapphire substrate. This sample was prepared by mechanical exfoliation of high purity $h$BN single crystals and transferred onto a 40 nm thick VO$_2$ film deposited on 500 $\mu$m thick sapphire by pulsed laser deposition. To determine the thickness of this flake at each position, we employed atomic force microscopy (AFM). Figure 2b shows an AFM image of the $h$BN flake shown in Figure 2a. The red dot in Figure 2a, 2b indicates the location of the quantum emitters 'A' and 'B' with emission wavelength of 600 nm and 620 nm, respectively. Figure 2c show AFM height profile across the lines (S-E) indicated in Figure 2b where flake thickness varies in the range 230-420 nm. At the location of the emitter 'A' and 'B' height of the flake is 310 nm and 340 nm respectively.

To locate the quantum emitters precisely, we performed confocal photoluminescence (PL) mapping in mode by which the sample was scanned point by point. Figure 2d shows a PL map over an area of 20 x 20 $\mu m^2$ on the $h$BN flake. The location of the quantum emitters 'A' and 'B' are highlighted by dashed circles. The single photon emission nature of these quantum emitters is evident from their second-order autocorrelation measurements indicating $g^2(0) < 0.5$ (see Figs. 3 a,b). Figures 2e, 2f show the PL spectra of these quantum emitters obtained for insulating and metallic VO₂. The emission spectra of each quantum emitter consist of a pronounced ZPL accompanied by a weaker phonon assisted emission. An increase in PL intensity at obtained for metallic VO₂ compared to insulating VO₂ is noticeable for all quantum emitters, which is indicative of a higher photon emission rate i.e. decrease in emission lifetime. This enhancement of emission rate is due to modification in LDOS owing to change in complex refractive index of VO₂ when switched from the insulating to metallic state. Recent experiment has reported that the decay rate of $h$BN quantum emitters remains constant even when heated up to 800 K (*42*) which further corroborates that the enhancement of decay rate is due to modification in LDOS rather than a thermal effect. The defect-based quantum emitter's dimensions are atomic scale, and thus lateral emitter size and location measurement is constrained by the optical diffraction limit (Supplementary Fig. S3).

To investigate the single-photon emission characteristics and decay lifetime of the quantum emitters, we measured their second-order intensity correlation functions $g^2(\tau)$ in both insulating and metallic phases of VO₂. In order to reduce the influence of background signal and noise, we corrected the raw $g^2_{raw}(\tau)$ using the function $g^2(\tau) = [g^2_{raw}(\tau) - (1 - \rho^2)]/\rho^2$, where $\rho = S/(S + B)$ where $S$ and $B$ refer to the signal and the background counts, respectively. This background corrected $g^2(\tau)$ was fitted with double exponential of the form (*20*)

$$g^2(\tau) = 1 - \rho^2\big[(1 + \zeta)e^{-\gamma_1|\tau|} - \zeta e^{-\gamma_2|\tau|}\big] \qquad (1)$$

where, $\zeta, \rho, \gamma_{1,2}$ are laser power-dependent parameters (*20,48*). Here, $\gamma_1$ and $\gamma_2$ are the faster and the slower decay time constants, respectively, for a three-level system. The second-order intensity correlation functions $g^2(\tau)$ under continuous wave excitation pumping for the quantum emitters 'A' and 'B' is shown in Figures 3a and 3b respectively, when VO₂ is insulating (blue dots) and metallic phase (red dots). The data for the metallic phase VO$_2$ configuration has been offset vertically for visual clarity. Equal-time coincidence counts $g^2(0)$ for each quantum emitter is less than 0.5, which indicates the presence of a single emitter. All measurements were performed at a constant 50 μW pump laser power which is orders of magnitude smaller compared to the saturation power of ∼ mW for $h$BN quantum emitters (*10,18,19*). Given the Fresnel reflections from all interfaces, which were analyzed using full-wave simulations, the excitation power within the flake along the axial direction is position dependent. From fitting our experimental data of correlation functions $g^2(\tau)$, we extracted the decay constants ($\gamma_{1,2}$) which has contributions from the spontaneous decay rates and the pump rate (*43,44*). The spontaneous decay rates ($\gamma$) of the emitters 'A' and 'B' are shown in the Table T1 (see supplementary information section S2 for details). The correlation functions over long-time scales are shown in Fig. S10. With an excitation power of 50 μW, the pump rates are ∼ 25-fold and 73-fold slower than the spontaneous decay rates for emitters 'A' and 'B', respectively, and thus make negligible contributions to the decay constants. From Table T1, we clearly see that for all the emitters, the decay rates are higher in the presence of a metallic-VO$_2$ when compared to an insulating-VO$_2$ configuration. Thus, the emitters 'A' and 'B' are located at distances, from the surface of VO$_2$, such that their optical environment is modified when VO$_2$ undergoes an insulator-to-metal transition.

To model the distance-dependence of the quantum emitter lifetime on VO$_2$ phase, we define the ratio of their decay rates in the insulating and metallic phases as $\beta$. Figures 3c, 3d show the two-dimensional plot of relative decay rate $\beta$ as a function of distance $d$ from the $h$BN/VO$_2$ interface and the polar orientation angle $\theta$ of the dipole for each quantum emitters 'A' and 'B' respectively. Each plot has three contour lines; a dashed line for the relative decay rates $\beta$ while the upper and the lower solid contour lines corresponds to the error in decay rate ( $\pm \Delta\beta$ ). Using the experimental values of $\gamma$ from the Table T1, we obtained the relative decay rates $\beta$ for the quantum emitters 'A' and 'B', as $0.818 \pm 0.108$ and $0.800 \pm 0.124$ respectively. From these simulations and the experimental values of $\beta$, it is evident that the quantum emitters are located within a narrow region at a distance $d \sim 21$ nm from the surface of $h$BN/VO$_2$ interface. However, the uncertainty in the axial position depends on the emitters' polar angle $\theta$. For emitter 'A', uncertainty (full width) varies from ∼ 13 nm at $\theta = 0^0$ to ∼ 22 nm at $\theta = 90^0$. Similarly, for quantum emitter 'B' uncertainty in their axial location varies from ∼ 15 nm to ∼ 23 nm at $\theta = 0^0$ and $\theta = 90^0$ respectively.

Next, we focus on emission polarimetry of the quantum emitters. Previous studies (*45*) have shown that the three-dimensional orientation $(\theta, \varphi)$ of a dipole can be directly extracted by analyzing polarization characteristics of its emitted light. Figures 4a, 4b is the emission polarization measurement from the quantum emitters 'A' and 'B' respectively. The data is fitted by the function (*45*)

$$I(\alpha) = I_{min} + (I_{max} - I_{min}) \cos^2(\alpha - \varphi) \qquad (2)$$

where $I_{min, \ max}$ and $\varphi$ are the fitting parameters. From the fit, we obtained for emitter 'A': $I_{min} = 0.356 \pm 0.013$, $I_{max} = 0.966 \pm 0.024$ and $\varphi = 175.7^0 \pm 1.0^0$. Similarly, for emitter 'B': $I_{min} = 0.318 \pm 0.034$, $I_{max} = 0.888 \pm 0.066$ and $\varphi = 109.9^0 \pm 2.9^0$. In emission polarimetry, the polar angle $\theta$ can be extracted from the degree of polarization of the emission defined as

$$\delta(\theta) = \frac{I_{max} - I_{min}}{I_{max} + I_{min}} \qquad (3)$$

From the fitting parameters $(I_{max}, I_{min})$, we obtained $\delta = 0.461 \pm 0.023$ and $\delta = 0.473 \pm 0.070$ for emitter 'A' and 'B' respectively. Figures 4c, 4d shows the calculated degree of polarization $\delta$ as a function of the polar orientation angle $\theta$ using the experimental values of numerical aperture (0.9), the refractive indices of $h$BN, VO$_2$ (insulating phase) and sapphire. The distance $d$ of the quantum emitters 'A' and 'B' from VO$_2$/sapphire substrate was set to $d \sim 20$ nm. From Figures 4(e,f) we clearly see that variation in the distance $d$ is negligible (dashed line). The red dot in Figs. 4e, 4f represents the measured value of $\delta$ and we extract the polar orientation angle $\theta = 20.5^0 \pm 3.6^0$ and $\theta = 21.2^0 \pm 4.5^0$ for emitter 'A' and 'B' respectively. In estimating the value of error in $\theta$ we accounted for the error in location $d$, which is shown Figs. 4(e,f) by solid lines. Figure 4(e) represents nanometer-scale axial location of emitter A with an uncertainty (full-width) of $\sim 15$ nm, oriented along $(\theta, \varphi) = (20.5^0 \pm 3.6^0, 175.7^0 \pm 1.0^0)$. Similarly, Figure 4(f) represents nanometer-scale axial location of emitter B with an uncertainty full-width of $\sim 17$ nm, oriented along $(\theta, \varphi) = (21.2^0 \pm 4.5^0, 109.9^0 \pm 2.9^0)$. The strong vertical component of the dipole orientation for both emitters found in the vicinity of $h$BN/VO$_2$ interface are in agreement with our simulations results shown in Figure S6. The predominant out of plane dipole orientation of these emitters can be a result of an out of plane atomic structure of the emitter as discussed in the literature. We believe the information provided by these measurements are of fundamental value in determining the underlying atomic structure of emitters.

Next, we investigated the feasibility of transferring the $h$BN flake from the $VO_2$ measurement substrate to a device substrate, which is a necessary capability for integration of quantum emitters with chip-based photonic components and waveguide circuits. In contrast to wet chemical transfer method (*46*), we utilized a polymer-assisted hot pickup technique (*30, 47,48*) to transfer the emitter-host $h$BN flake from $VO_2$ to a receiving $SiO_2$/Si device substrate (Supplementary Figure S13). Figure 5a, 5b show the optical image of the flake before and after the transfer. We were able to relocate the emitters 'A' and 'B' on the device substrate by performing confocal PL mapping (Figure 5c) and matching the spectral (Figure 5d) and spatial signatures of both emitters before and after the transfer process.

To summarize, we have demonstrated an experimental technique by which the axial position of quantum emitters in a multi-layered $h$BN flake can be extracted with nanometer-scale accuracy by exploiting the modification of photonic density of states using a phase change optical material, $VO_2$. Here, we tailor the optical environment of an emitter in the vicinity of $VO_2$/Sapphire substrate which generates a sharply distance-dependent PL response. By performing time-resolved fluorescence spectroscopy, supplemented with emission polarimetry, several specific quantum emitters were identified at an axial distance of $\sim 21$ nm from the $h$BN/$VO_2$ interface while also determining their full dipolar orientation ($\theta$, $\varphi$). Furthermore, we utilized a polymer-assisted hot pickup technique to transfer the identified $h$BN emitters from a $VO_2$ measurement substrate to a $SiO_2$/Si device substrate, which opens the door to coupling of fully characterized emitters, where each emitter has undergone precise measurement of axial position and orientation. It is worth noting that any phase-change material which experiences a sharp change in optical properties would be suitable for this purpose. However, $VO_2$ is particularly interesting because its insulator-to-metal transition happens near room temperature and is thus well suited to dynamically control emission rates of quantum emitters (*49,50*) near room temperature. Owing to the broadband nature of change in the dielectric function of $VO_2$ when switched from the insulating to metallic phase, our technique could also be extended to locating other visible or infrared quantum emitters (*10*).

## Acknowledgments

We thank S. Nam for the useful discussions**.** This work was supported by the DOE "Photonics at Thermodynamic Limits" Energy Frontier Research Center under grant DE-SC0019140 and by the Boeing Company.

## Author contributions:

P. K. J, H. A, Y. K, and H. A. A conceived and developed the idea. P. K. J prepared the *h*BN flakes and performed AFM measurements. H. A performed the optical characterization of *h*BN flakes and correlation measurements. P. K. J and H. A performed emission polarimetry of *h*BN quantum emitters. Y. K prepared $VO_2$/Sapphire sample; performed ellipsometry, full-wave simulations, optical, and AFM characterizations of $VO_2$ thin films. S. B developed and performed transfer of $h$BN flake from $VO_2$/Sapphire substrate onto $SiO_2$/Si for further studies. P. K. J simulated the optical response of $h$BN quantum emitters and theoretical estimations with inputs from all co-authors. H.A.A. supervised all the experiments, calculations, and data collection. All authors contributed to the data interpretation, presentation, and writing of the manuscript.

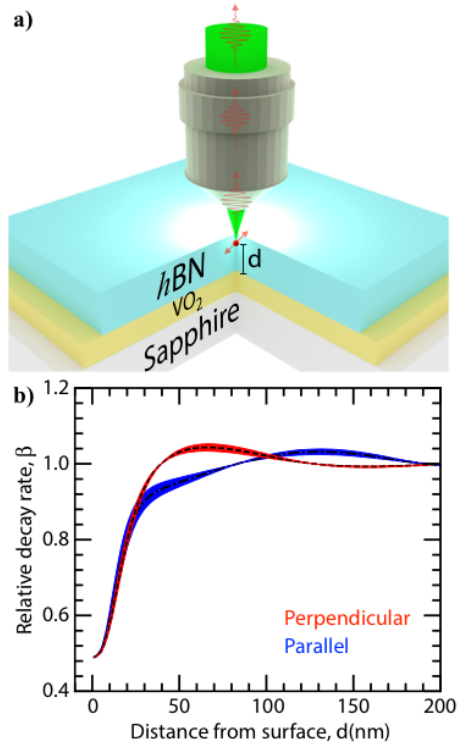

**Figure 1:** Experimental schematic and distance-dependent modulation of relative decay rates. (a) Schematic of a quantum emitter in an atomically thin crystal of hexagonal boron nitride (*h*BN) located within the thickness of a flake on a substrate which consists of a thin layer of vanadium dioxide (VO$_2$) deposited on a sapphire crystal. (b) Relative decay rate $\beta = \gamma_{Insulating}/\gamma_{Metallic}$ as a function of distance *d* of a quantum emitter from the surface of VO$_2$ when switched from the insulating to metallic state. The blue and red curves refer to quantum emitters oriented parallel and perpendicular to the surface respectively and the shaded regions corresponds to the typical quantum yield (QY) range of 0.6-1.0 of *h*BN quantum emitters with zero-phonon line around 600 nm (*46*). For numerical simulation we considered the emission wavelength of 600 nm for the quantum emitter corresponding to the emitter 'A' (Fig. 2e). The dashed line within the shaded region corresponds to QY = 0.79 as estimated from our experimental data (Supplementary Sec. S2). The refractive indices of the upper medium, *h*BN, and that of sapphire were set to 1, 1.82, and 1.77 respectively. The refractive index of VO$_2$ at 600 nm was set to 3.05 + 0.42i (insulating state) and 2.57 + 0.64i (metallic state) from our ellipsometric data (Supplementary Fig. S1).

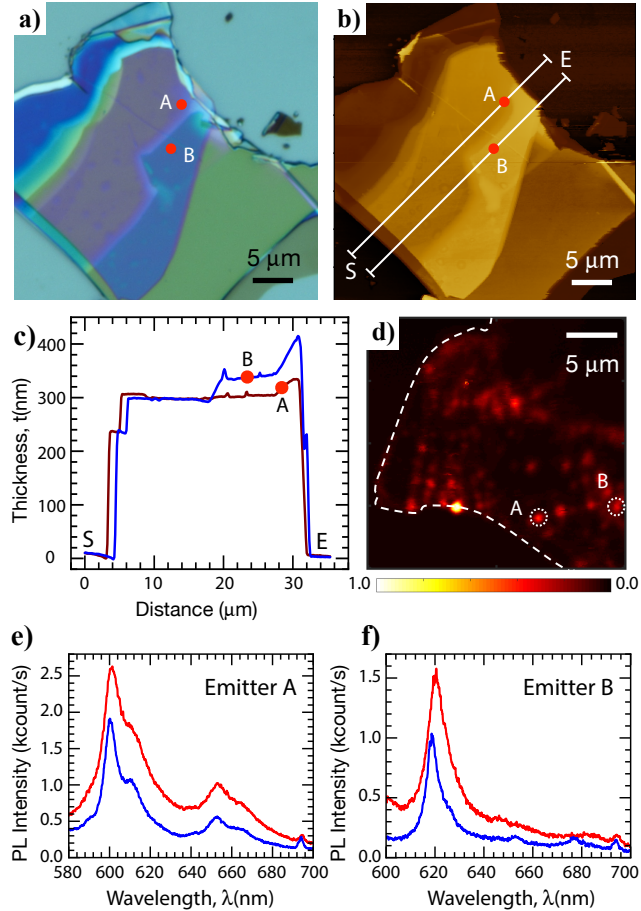

**Figure 2:** Characterization of the exfoliated flake and spectra of *h*BN quantum emitters. (a) Optical image of the mechanically exfoliated *h*BN flake on VO$_2$/Sapphire substrate. (b) Atomic force microscopy image of the flake shown in (a). The red dots on the traces (S-E) in (b) and (a) indicate the position of emitters 'A', and 'B' with emission wavelength of 600 nm and 620 nm, respectively. (c) Line profiles along the region indicated by the trace in (b). At the location of the emitter 'A' and 'B' height of the flake is 310 nm and 340 nm respectively. (d) Confocal photoluminescence (PL) map of the *h*BN flake. The position of two single photon emitters is marked by white circles. The edge of the flake in marked by white dashed line. The PL spectra of emitters 'A' and 'B' shown in (e) and (f) respectively were obtained with VO$_2$ in insulating state (blue) and metallic state (red).

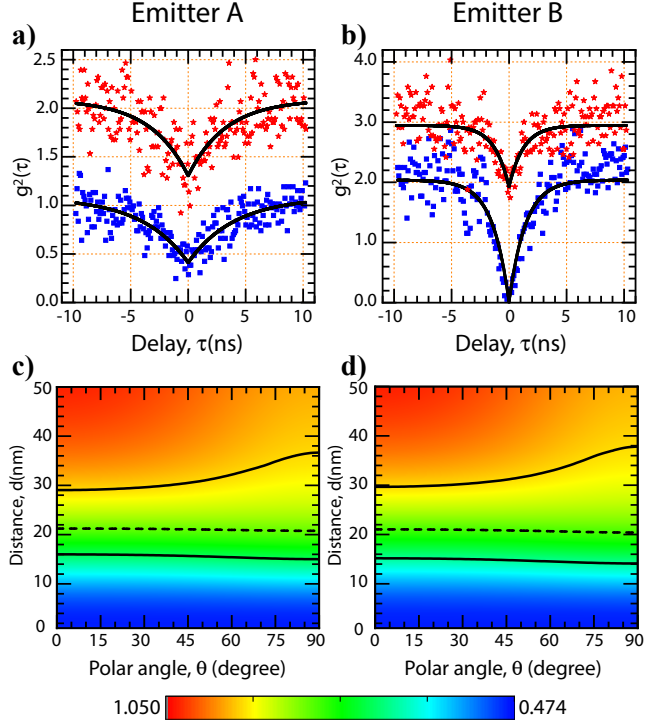

**Figure 3:** Single photon source characterization and axial location in *h*BN flake. Plot of the second order photon correlation measurement, g²(τ) for the emitters 'A' and 'B', in (a) and (b) respectively. The experimental data, blue squares for insulating VO₂ phase and red stars for metallic VO₂, were fitted using Eq. (1) to obtain the decay rates of the emitters. From the fit, we calculated the relative decay rates $\beta = \gamma_{Insulating}/\gamma_{Metallic}$ for the three emitters 'A' and 'B' as $0.818 \pm 0.108$ and $0.800 \pm 0.124$ respectively. For clarity, g²(τ) data obtained for metallic VO₂ in (a), and (b) were shifted by 1 and 1.5 respectively. Plot of the relative decay rates $\beta$ as a function of the distance (*d*) from the surface of VO₂ and the polar angle (θ) for the emitters 'A' and 'B' are shown in (c) and (d) respectively. The dashed contour lines in (c) and (d) corresponds to the experimental value of $\beta$ obtained from (a) and (b) respectively, while the solid lines correspond to the error ($\pm\Delta\beta$) in the ratio.

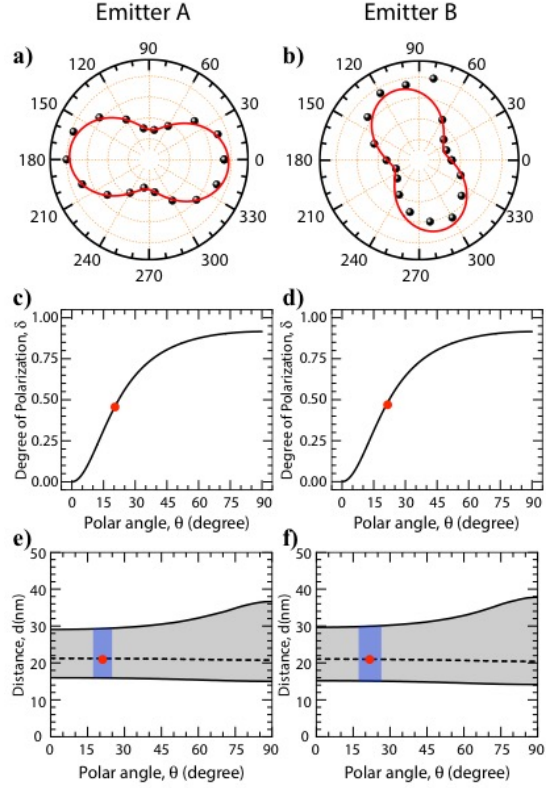

**Figure 4:** 3D-Orientation of *h*BN quantum emitters and nanometric axial location. (a,b) Polar plots of the photoluminescence (PL) intensity of the emitter 'A' and emitter 'B' respectively as a function of the emission polarization analysis angle ($\alpha$). The PL data (solid spheres) were fitted using Eq. (2) to extract the azimuthal angle ($\varphi$) of the emitters and the degree of polarization ($\delta$). From the fit, we deduce that for emitter A, $\varphi$ = 175.7° ± 1.0°; $\delta$ = 0.461 ± 0.023 and for emitter B, $\varphi$ = 109.9° ± 2.9°; $\delta$ = 0.473 ± 0.070. (c, d) Calculated value of degree of polarization ($\delta$) as a function of the polar angle ($\theta$) of the emitters located at a distance of ~ 21 nm from the surface of VO$_2$. The red dots in (c) and (d) corresponds to the experimental value of $\delta$ obtained from (a) and (b) respectively. The extracted value of the polar angle for emitter 'A' and 'B' are $\theta$ = 20.5° ± 3.6° and $\theta$ = 21.2° ± 4.5° respectively. (e, f) Purple shaded region shows the range of the distance (*d*) and the polar angle ($\theta$) of the emitters 'A' and 'B' respectively based on our experimental data and simulations.

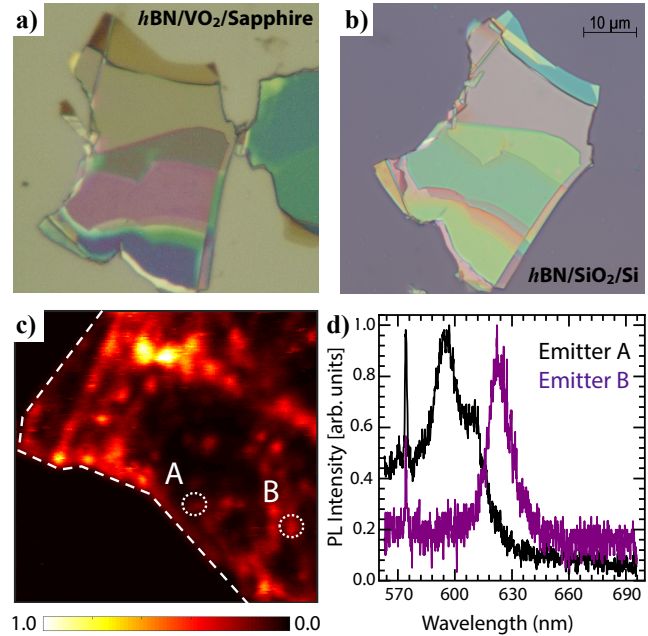

**Figure 5:** (a,b) Optical image of the flake on VO₂/Sapphire substrate (before transfer) and SiO₂/Si substrate (after transfer) respectively. (c) Confocal photoluminescence (PL) map of the flake on SiO₂/Si substrate. Encircled (dashed while circle) highlights the region where emitters 'A' and 'B' are located. (d) PL spectra from the region highlighted in (c) shows the presence of emitters 'A' and 'B' which matched with the spectra obtained from the same location before transfer. The sharp line ~ 575 nm corresponds to Raman line of $h$BN.

# Supplementary Materials for

**Nanoscale axial position and orientation measurement of hexagonal boron nitride quantum emitters using a tunable nanophotonic environment**

Pankaj K. Jha [†], Hamidreza Akbari[†], Yonghwi Kim[†], Souvik Biswas, Harry A. Atwater[*]

[†]These authors contributed equally to this work.

[*]Corresponding author: Harry A. Atwater (haa@caltech.edu)

This PDF file includes:

## Section S1: VO₂ characterization, optical properties, and optimization

We performed visible ellipsometry measurement to extract real and imaginary parts of refractive index of VO₂ at 30°C and 100°C. We clearly observe the switching of the optical constants of the VO₂ film when reaching the temperature triggered phase transition. It is worth noting that, $n$ decreases in the wavelength range, while $k$ increases as VO₂ undergoes an insulator-to-metal transition. To date, quantum emitters in $h$BN exhibit ZPL emission in the range 550-800 nm; thus, VO₂ is well-suited to modify the optical environment of $h$BN-based quantum emitters.

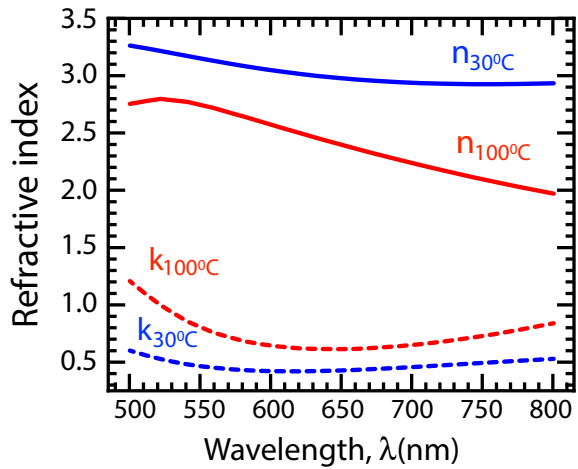

Fig. S1: Complex refractive index of VO2 in insulating and metallic phase. Extracted value of real (n) and imaginary (k) part of the complex refractive index of 40 nm thick layer of VO2 deposited on sapphire crystal at 30°C (blue) and 100°C (red) from our ellipsometric data.

We also measured the response of the Peltier element to measure how long it takes to reach the stabilized temperature (Fig. S2). From room temperature to 100°C it takes less than 100 seconds but from 100°C to room temperature it takes around 1000 seconds, hence each time we waited 5 minutes when we were heating up and waited an hour when we were cooling down.

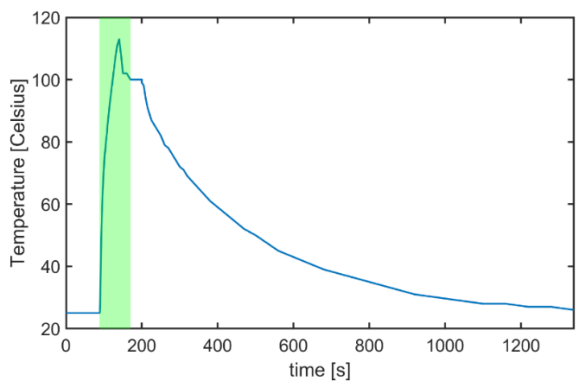

Fig. S2. Temporal response of Peltier stage. Response of the Peltier stage as a function of time for ramp up and cooldown. Shaded region shows the ramp up region and the temperature is set to 100°C for 30 seconds and then the temperature decreases as a result of convection.

We employed the Bruggeman effective medium approximation to estimate the intermediate optical constants as a function of volume fraction of the metallic phase of the VO$_2$ layer (*53*). The legend R (0.00) denotes the purely insulating phase VO$_2$, while the legend R(1.00) represents the purely metallic phase VO$_2$. The estimated optical constants were used in full-wave simulations to monitor the reflectance values at normal incidence. The full-wave simulation results is consistent with the measured spectra, which validates the refractive index and the thickness of the VO$_2$ film that we characterized.

In order to characterize the optical properties of the VO$_2$ film grown on a sapphire substrate, we measured reflectance spectra in the visible range. The temperature-dependent reflectance curves were measured by using a microscope spectrometer with an external heating stage. Figure S3 shows the temperature-dependent reflectance modulation in the 30–100°C range of the 40-nm–thick VO$_2$ film for the heating cycle. We observed a gradual reflectance change as a result of the insulator-to-metal transition when we slowly changed the substrate temperature. The temperature-dependent curves represent a gradual decrease of the reflectance as the VO$_2$ film becomes a lower index and a more lossy metallic state.

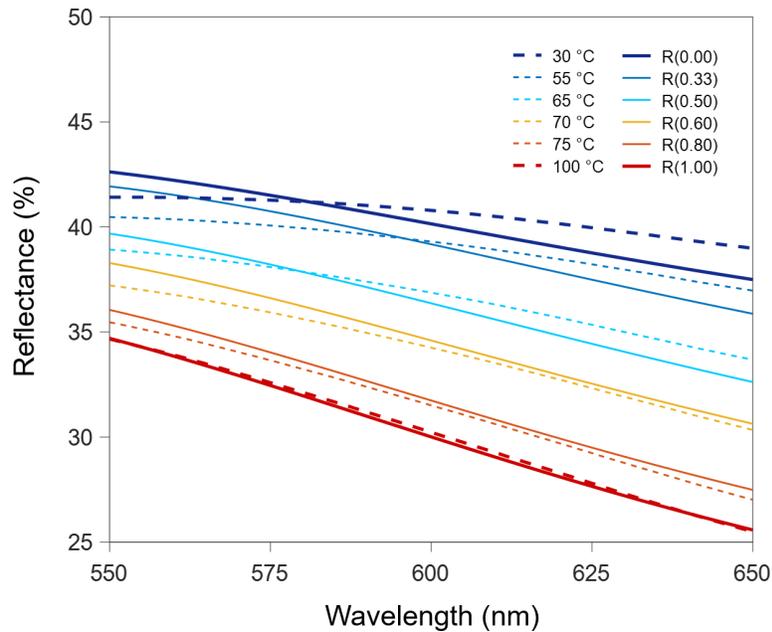

Fig. S3. Temperature-dependent reflectance spectra of a VO$_2$ film during the heating cycle (dotted lines) and simulated reflectance curves of the film using the Bruggeman effective medium theory (solid lines). We observe gradual reflectance variations upon phase transition in VO$_2$, where the volume fraction of metallic phase VO$_2$ gradually evolves within the insulating phase VO$_2$ host.

Numerical simulations were performed using the commercial finite element software COMSOL Multiphysics for frequency domain electromagnetic full-field calculations. The simulation domain was truncated using perfectly matched layers. We modelled a quantum emitter as an electric point dipole source and calculated the spontaneous emission rate by probing the field at the location of the dipole. The non-radiative contribution to the decay rate was calculated by probing the heat generated in the lossy material (VO$_2$). The complex refractive index of VO$_2$ at the ZPL of emitters A, B, and C were taken from our ellipsometric data (Supplementary Fig. S1). The refractive index of hBN was set to be 1.82 at 600 nm, and a weak linear variation was employed to calculate refractive index at 570 nm and 620 nm as 1.84 and 1.81 respectively (*40, 51*). Sapphire has wavelength dependent refractive index (*52*).

We performed full-wave electromagnetic simulations using finite difference time domain methods to estimate the optimal thickness for a VO$_2$ film in order provide the largest optical contrast in the photoluminescence wavelength range. We used a commercial Lumerical finite-difference time-domain software package to obtain the reflectance spectra of the VO$_2$ film on a sapphire substrate. The reflectance maps of the thin film structure were monitored by varying the thickness of the VO$_2$ film for both insulating and metallic phases. Finally, the thickness of VO$_2$ film was optimized based on the FOM equation shown below. Fig. S4 represents the simulated FOM map, which indicates that about 40-nm-thick VO$_2$ film shows the largest reflectance contrast at the zero-phonon lines of our exfoliated hBN flake. Based on this analysis, we grew 40-nm-thick VO$_2$ films as explained in Section S8.

$$FOM = \frac{|R_{Metallic} - R_{Insulating}|}{R_{Metallic} + R_{Insulating}}$$

Here, $R_{Metallic}$: metallic phase reflectance and $R_{Insulating}$: insulating phase reflectance.

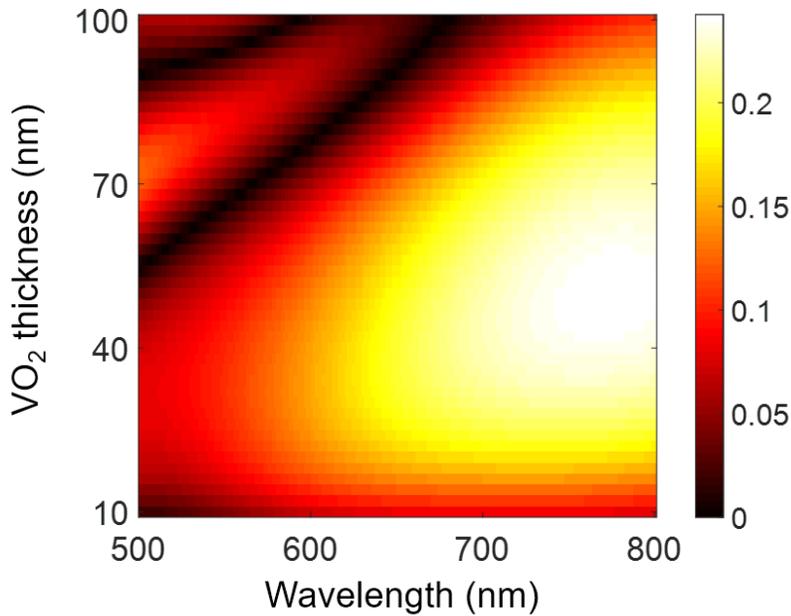

Fig. S4. Optimization of VO$_2$ thickness. Figure of merit map (FOM) of a VO$_2$ film on a sapphire substrate obtained by full-wave simulations. The FOM indicates the reflectance contrasts between the insulating and metallic phases of VO$_2$. The results indicate that the largest optical contrasts at the zero-phonon lines when VO$_2$ thickness is about 40 nm.

A 40-nm thick vanadium dioxide (VO$_2$) film was formed on a cleaned c-plane single side polished sapphire substrate by pulsed laser deposition. A high-power pulsed laser beam vaporizes a vanadium target which deposits a thin film on the sapphire substrates in the presence of 5 m-Torr oxygen gas at an elevated temperature (650°C). First, we confirmed that the surface of the VO$_2$ film grown on the sapphire substrate was uniform and continuous with root mean square roughness of around 1.5 nm that were measured by atomic force microscopy (Figure S5). The measurement result shows the morphology of the VO₂ film which consists of smooth and continuous small grains.

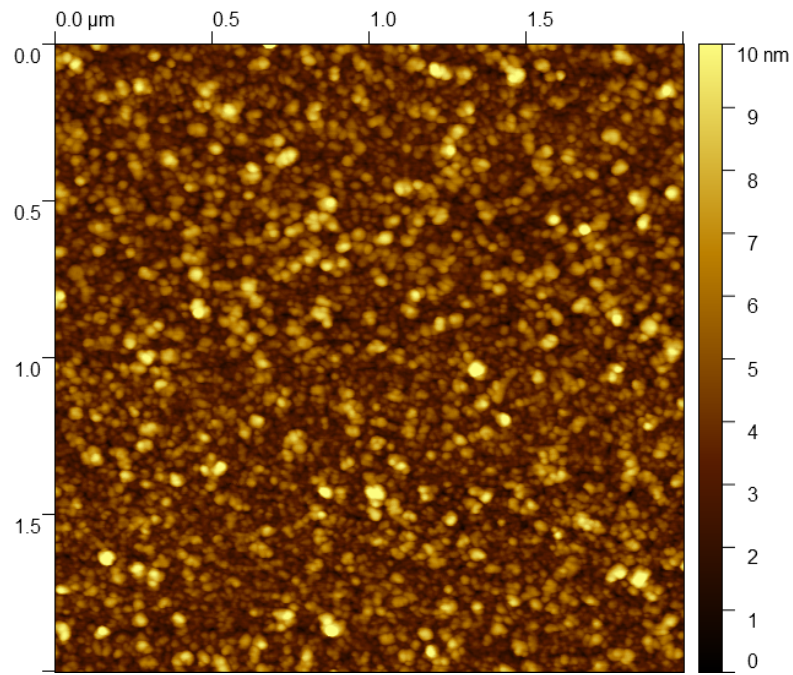

Fig. S5. Surface characterization of VO₂ film. Atomic force microscope measurement of VO₂ film deposited on a sapphire substrate. This result shows that the film is uniformly grown, and the root-mean-square roughness is around 1.5 nm (2 μm × 2 μm) which is much smaller than thickness of the film which is 40 nm.

## Section S2: LDOS engineering with phase change material

Figure S6 shows the individual contribution from radiative and non-radiative channels to the decay rates $\gamma_{Insulating}$ (Figs. S6 A,B) and $\gamma_{Metallic}$ (Figs. S6 C,D) for both parallel and perpendicular orientations as a function of the distance ($d$) from $h$BN/VO$_2$ interface. The non-radiative decay rate of the emitter corresponds to the emission of a photon which is absorbed in the lossy material VO$_2$. For numerical simulation we considered an $h$BN quantum emitter with ZPL wavelength of $\sim$ 600 nm and PLQY of 0.79 (Supplementary Section S4). Figure S6 illustrates that at distances in the range $\sim$ 0-15 nm, decay rate $\gamma$ is dominated by the non-radiative channel and thus the emitters located within this range of distances are "hidden" owing to a lack of light emission available for collection objective lens. However, the radiation contribution to the total decay rate is stronger for vertical dipole compared to the horizontal dipole within first $\sim$50 nm with both states of VO$_2$. Thus, emitters found in the vicinity of $h$BN/VO$_2$ interface have a higher probability of dipole moment oriented along the vertical direction.

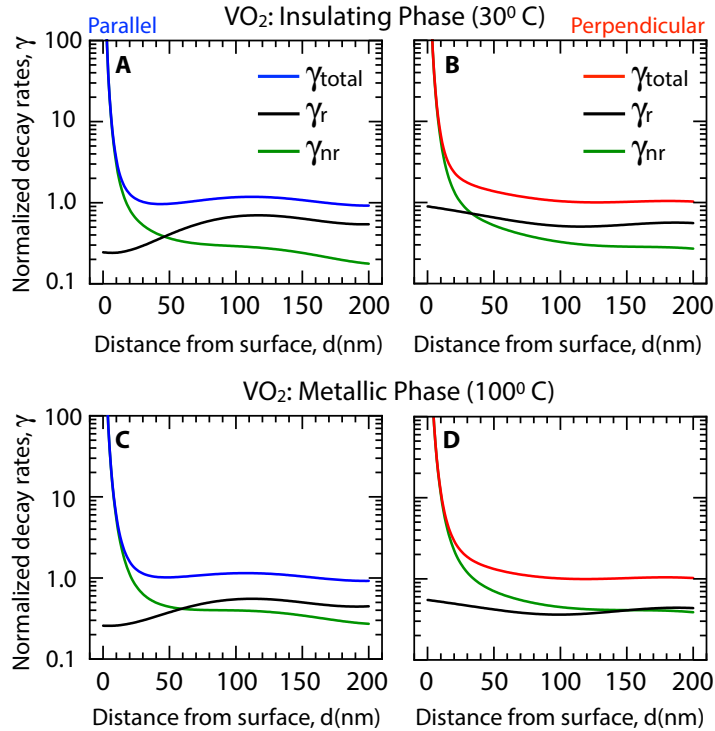

Figure S6. Radiative and non-radiative decay rates. Total ($\gamma_{total}/\gamma_0$), radiative ($\gamma_r/\gamma_0$), non-radiative ($\gamma_{nr}/\gamma_0$), rates of spontaneous emission of a quantum emitter as a function of distance from VO2 surface for two dipole orientation i.e. parallel ($\theta = 90°$) and perpendicular ($\theta = 0°$) to the surface. For numerical simulation we considered the emission wavelength $\lambda_0 = 600$ nm and the quantum yield QY = 0.79 corresponding to the emitter 'A'. (A, B) Insulating phase of VO2, (C, D) metallic phase of VO2. Free-space decay rate $\gamma_0 = n\omega_0^3|\varrho|^2/3\pi\varepsilon_0\hbar c^3$, with c being the speed of light, n the refractive index of hBN, $\omega_0$ the atomic transition frequency, $\hbar$ the reduced Planck's constant, and $\varrho$ the amplitude of the dipole moment vector. The refractive indices of the upper medium, hBN, and that of sapphire were set to 1, 1.82, and 1.77 respectively. The refractive index of VO2 at 600 nm was set to 3.05 + 0.42i (insulating phase) and 2.57 + 0.64i (metallic phase) from our ellipsometric data (Supplementary Fig. S1).

## Section S3: Optical characterization of emitters

Optical characterization of samples was performed in a home-built confocal microscope capable of optical spectroscopy in visible range (Princeton HRS 300 system) and intensity auto-correlation measurement ($g_2(\tau)$) in a Hanbury Brown Twiss (HBT) configuration using a 50-50 beam splitter and two avalanche photo diodes (model). We used a fast scanning mirror (Newport) and a $4f$ telecentric configuration to perform photoluminescence mapping. The microscope uses a 532 nm CW laser (Cobolt) in order to pump emitters in $h$BN and a 100X objective (Leica) to focus the beam on the sample and used 50 µW power of laser (before objective) for all emitters. A quarter wave plate was put in the beam path at $45^0$ orientation with respect to linear polarization of laser in order to produce circularly polarized light. We pumped with circularly polarized light to excite all emitters irrespective of their in-plane dipole orientation. A tunable bandpass filter (Semrock versachrome) was used to only pass the zero-phonon line on the emitter into HBT setup to reduce background. Schematic of our optical characterization setup can be seen in Fig. S7.

Annealing in an inert environment is routinely used to create or activate quantum emitters in diamond as well as $h$BN. For the $h$BN samples described here, we annealed a bulk crystal of $h$BN at 950 °C in a 1 bar pressure argon gas for 30 minutes before exfoliation. We mounted our sample consisting of $h$BN/VO$_2$/sapphire on a Peltier heating stage.

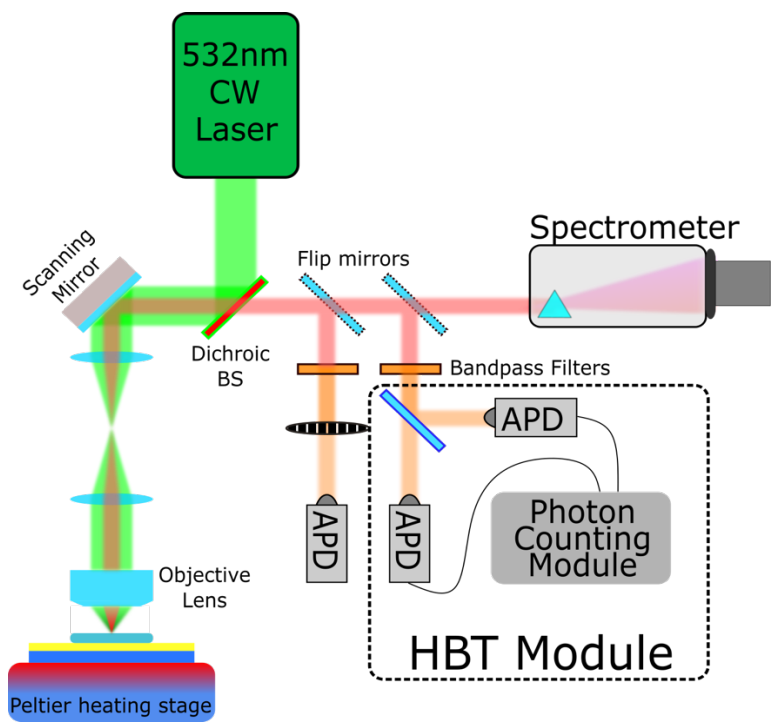

Fig. S7. Experimental setup. Schematic of the homebuilt confocal microscope used to characterize hBN quantum emitters and perform correlation measurements.

To perform background subtraction for g⁽²⁾ measurements we scanned a small PL map (5umX5um) around each emitter in the same optical condition as the g⁽²⁾ measurement was performed (i.e. using bandpass filter to only pass through the zero phonon line). Then by fitting a gaussian profile to the spatial photoluminescence we extract the value of background (B) and signal(S), then we define parameter ρ as $\rho = S/(S + B)$. If the raw (i.e. background is not corrected) g⁽²⁾ function is $g_{raw}^{(2)}$ then the background corrected intensity autocorrelation $g_{bgc}^{(2)}$ is calculated as:

$$g_{bgc}^{(2)}(t) = \frac{g_{raw}^{(2)}(t) - (1 - \rho^2)}{\rho^2}$$

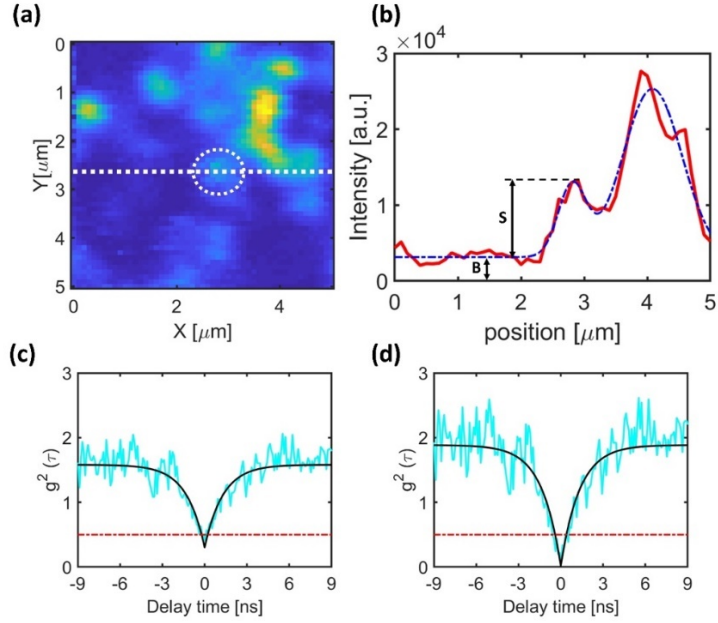

Fig. S8. Background subtraction for correlation measurements. Background subtraction for $g^2(\tau)$ measurement of emitter B. (a) shows a PL map around the emitter, (b) shows a profile of the PL map shown by a dashed line in (a) with a gaussian fit to it, and values for signal and background. (c) shows $g^2(\tau)$ result before background subtraction and (d) shows $g^2(\tau)$ result after background subtraction.

A similar method is used for background subtraction of polarization measurement. We subtracted ½ of the background value extracted from the fit mentioned above from the angle dependent polarization.

We also studied a bright emitter with ZPL at 570 nm however before performing polarization measurement this emitter bleached and we couldn't perform a complete analysis similar to emitters 'A' and 'B'.

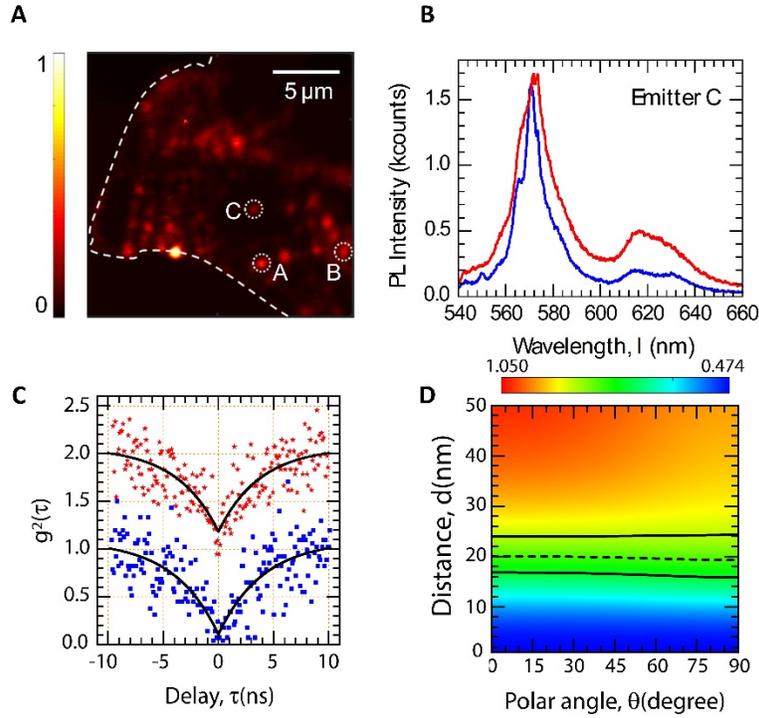

Fig. S9. Emitter C. In (A) we can see the position of this emitter in the PL map and spectra (B) and autocorrelation measurement result(C) at 30C and 100C. By comparing the ratio of decay rates to simulations at 570 nm we extracted the axial position of this emitter as can be seen in (D).

The full information of decay rates of three emitters under study can be seen in table S1.

| Emitter | Decay rate, $\gamma_{\text{Insulating}}$ (MHz) | Decay rate, $\gamma_{\text{Metallic}}$ (MHz) |
|---------|------------------------------------------------|----------------------------------------------|
| A | $249 \pm 15$ | $304 \pm 33$ |
| B | $693 \pm 45$ | $865 \pm 122$ |
| C | $268 \pm 16$ | $334 \pm 18$ |

Table S1. Spontaneous emission rates of emitters. Spontaneous decay rates of the emitters 'A', 'B', and 'C' in the vicinity of VO$_2$ in insulating phase and metallic phase. The decay rates are estimated based on the fitting of g$_2(\tau)$ and subtracting the contribution of pump-dependent excitation rate (see Supplementary section S2 for details).

Microsecond scale autocorrelation function of three emitters at both 30ºC and 100ºC is plotted in figure S10 to demonstrate how g$^{(2)}$ of all three emitters at long time scales approaches 1.

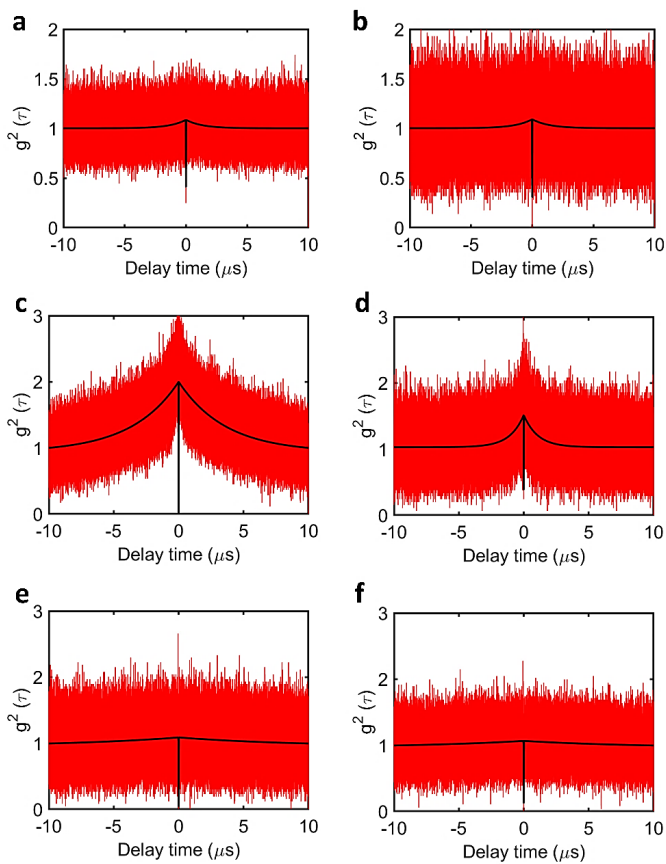

Fig. S10. Full-scale correlation data for the emitters A, B, C. Plot of the correlation measurement for Emitters 'A' (a,b), Emitter 'B' (c,d) and Emitter 'C' (e,f) as a function of delay over the range of microseconds. The data in (a,c,e) and (b,d,f) are at room temperature and 100°C respectively.

## Section S4: Photophysics of *h*BN quantum emitters

We model single *h*BN quantum emitter as three-level system that consists of a ground state $|b\rangle$, excited state $|a\rangle$ and metastable state $|c\rangle$ as shown in Fig. S11. The emission characteristics of this quantum emitter can be described by rate equations for population of the three levels (*43*).

$$\dot{\varrho}_{aa} = R\varrho_{bb} - (\gamma_b + \gamma_a)\varrho_{aa} \qquad (S1)$$
$$\dot{\varrho}_{cc} = \gamma_a\varrho_{aa} - \gamma_c\varrho_{cc} \qquad (S2)$$

Here, $R$ is the rate of excitation and $\gamma_i$ (with $i= a,\ b,c$) is the decay rate of population (radiative and non-radiative combined).

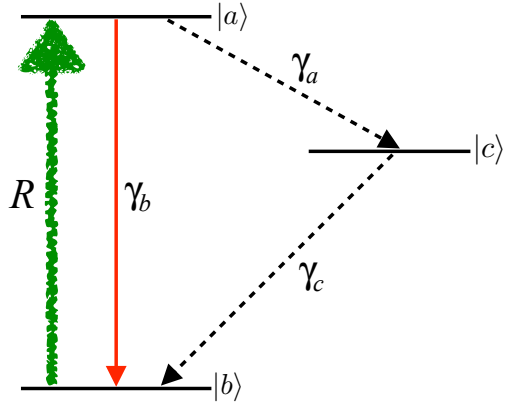

Fig. S11: Three-level system with corresponding decay and excitation rates.

These equations are also supplemented with population conservation equation $\varrho_{aa} + \varrho_{bb} + \varrho_{cc} = 1$. Using the rate equations, one can derive the analytical expression for the normalized "ideal" second-order autocorrelation function $g^2(\tau)$ as:

$$g^2(\tau) = 1 - \left[(1+\zeta)e^{-\lambda_1|t|} - \zeta e^{-\lambda_2|t|}\right] \qquad (S3)$$

where the coefficients are given in the limit ($\gamma_b \gg \gamma_a, \gamma_c$)

$$\gamma_1 = R + \gamma_b \qquad (S4)$$
$$\gamma_2 = \gamma_c + \frac{R\gamma_a}{(R+\gamma_b)} \qquad (S5)$$
$$\zeta = \frac{R\gamma_a}{[\gamma_c(R+\gamma_b)]} \qquad (S6)$$

However, in the presence of a background such as laser scatter or diffused photoluminescence Eq. (S3) takes a modified form (*20*)

$$g^2(\tau) = 1 - \rho^2\left[(1+\zeta)e^{-\gamma_1|\tau|} - \zeta e^{-\gamma_2|\tau|}\right] \qquad (S7)$$

where, $\rho$ quantifies the background and $g^2(0) = 1 - \rho^2$. The decay rate of the excited state $|a\rangle$ is $\gamma_t = \gamma_a + \gamma_b \approx \gamma_b$ as the transition rate to the metastable state $|c\rangle$ is orders of magnitude smaller compared to the decay rate $\gamma_b$. From Eq. (S4) we can obtain the decay rate of the quantum emitter using the pump-power-dependent decay rate $\gamma_1$ as $\gamma = \gamma_1 - R$. In our experiment, the focal spot was near diffraction limit ($\sim 0.5\ \mu m$), the average pump power was kept constant at $P_{avg} = 50\ \mu W$ for all the measurements.

The spontaneous decay $\gamma$ and the absorption cross-section $\sigma$ can be estimated by measuring the pump-power-dependent decay rate $\gamma_1$. Fig. S12 (a) shows the evolution of the decay rate $\gamma_1$ as a function of excitation power at the location of the emitter after taking into account Fresnel reflections from all interfaces. From our full wave-simulation, for emitter A ($\lambda = 600\ nm$) the excitation power drops to $\tilde{P}_{avg} \cong 0.6\ P_{avg}$. From the linear fit, we obtained the spontaneous decay $\gamma = 249$ MHz by extrapolating the linear fit to zero excitation power. At excitation power $\tilde{P}_{avg} = 30\ \mu W$ the absorption rate $R = 9.5$ MHz. Thus, in our experiment, the contribution of the excitation rate $R$ is negligible compared to the spontaneous decay rate i.e. $\gamma \cong \gamma_1$. The absorption rate is given by $R = \sigma\ I/h\nu$ (44) with absorption cross-section $\sigma$, excitation intensity $I$, and excitation frequency $\nu$. Using the absorption rate $R = 9.5$ MHz at excitation power $\tilde{P}_{avg} = 30\ \mu W$ and a nearly diffraction limited focal spot, we estimated the absorption cross-section $\sigma \cong 9.28\ \text{x}\ 10^{-16}\ cm^2$. Similar absorption cross-section for $h$BN emitters have been reported (54). Next, we estimated the intersystem crossing rates by fitting $\gamma_2$ with Eq. (S5). Here, we treated $\gamma_a$ as constant (44) and $\gamma_c$ as excitation power-dependent rate of the form $\gamma_c = \gamma_c^0 + \frac{uP}{P+v}$. From our fitting we obtained $\gamma_a = 5.84$ MHz, $\gamma_c^0 = 0.40$ MHz, $u = 0.89$ MHz, $v = 0.05$ mW.

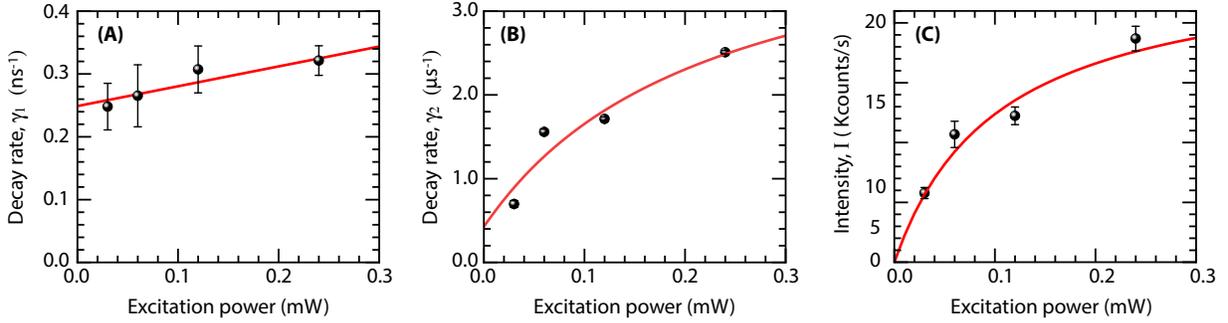

Fig. S12: Plot of the decay rates $\boldsymbol{\gamma_{1,2}}$ as a function of the excitation power at the location of the emitter. We fitted the experimental data shown in (A, B) with Eqs. (S4, S5) respectively. (C) Plot of the intensity as a function of excitation power to determine the saturation count rates ($I_{sat}$) and saturation power ($P_{sat}$).

Internal quantum yield in an important parameter for a quantum emitter which can be estimated from the value of rate coefficients and the maximum single photon rates i.e. the saturation count rate $I_\infty$ as (44)

$$I_\infty = \eta_c \eta_q\ \frac{\gamma_b}{1 + \gamma_a(\gamma_c^0 + u)^{-1}} \qquad (S8)$$

Here, $\eta_c$ is the collection efficiency of the experimental set up, $\eta_q$ is the quantum efficiency, $\gamma_b, \gamma_a, \gamma_c^0, u$ are rate coefficients which can be obtained from power dependent de-shelving model. We fitted the power-dependent PL intensity and fitted with the function $I = I_{sat}\left(\frac{P}{P+P_{sat}}\right)$ and obtained $I_{sat} = 0.025$ MHz and $P_{sat} = 104\ \mu W$. From the rate coefficients, we obtain $\eta_c \eta_q = 5.63\ \text{x}\ 10^{-4}$. Taking into account the collection, transmission, and coupling efficiency of the optical element of the experimental set up, the collection efficiency $\eta_c = 8.87\ \text{x}\ 10^{-4}$ which gives $\eta_q = 0.634$. The quantum efficiency $\eta_q$ has contribution from the internal quantum yield, radiative Purcell enhancement, and the non-radiative loss due to ohmic loss in $VO_2$. Using simple algebra, we obtain

$$\frac{1}{\eta_q} = \frac{1}{\eta_l} + \frac{1}{\eta_e}\left(\frac{1}{\eta_i} - 1\right) \qquad (S9)$$

Here, $\eta_l$ accounts for the rate of energy loss in $VO_2$, $\eta_i$ is the internal quantum yield of the *h*BN emitter, and $\eta_e$ is the radiative Purcell enhancement with respect to the homogenous medium. From our full-wave simulation, we obtained $\eta_l = 0.76$ , $\eta_e = 0.324$ which yield $\eta_i = 0.793$.

# Section S5: Transfer of $h$BN flake from $VO_2$ to $SiO_2$ substrate.

After the completion of emitter localization in $h$BN on $VO_2$, the flake can be deterministically transferred on to a different, arbitrary substrate for further studies, such as integration into resonant photonic structures like waveguides, photonic crystal cavities, DBRs, etc. The host $h$BN flake was deterministically transferred from $VO_2$ on to $SiO_2$ using the polymer assisted hot-pickup technique. A transfer handle with a hemispherical PDMS block (to allow precise point contact and deterministic choice of flake) coated with a thin polycarbonate (PC) film was used. The flake was picked up from $VO_2$ between 130-140ºC (contact and pickup were both done at elevated temperatures to increase yield) and dropped onto $SiO_2$ at 200ºC, following which the PC film was removed by gentle chloroform and isopropanol treatment.

The polymer assisted hot pickup-based transfer technique enables an accuracy in spatial positioning on the desired substrate close to 1-2μm. In principle, it is limited by two factors – optical resolving power of the microscope used during the transfer and the micromanipulators used to position the flake with respect to the substrate. In our experiments, a 50x objective was used, which allowed us to identify feature sizes of about 1μm and a manual micromanipulator set was used which had a bidirectional repeatability of approximately 2μm. In addition, the desired substrate can be placed on a rotational stage and the host $h$BN flake can transferred with a rotational alignment precision of about 0.1 degrees. However, in our experiment the uncertainty in rotational alignment is limited by uncertainty in the in-plane dipole moment which is about 3-5 degrees. This enables high-accuracy angular alignment of the in-plane dipole moment of the $h$BN flake to the desired substrate containing any nanophotonic design sensitive to dipole orientation.

In order to facilitate that we fabricated "transfer handles" as follows:  A 1 cm x 1 cm x 0.5 cm block of cured PDMS was cut and put on a glass slide. Drops of uncured PDMS (Sylgard 10:1 ratio) were poured onto the block so that the PDMS assumes a hemispherical lens-like shape. This was cured overnight at 80C in an oven. The curvature of the PDMS ensured deterministic contact at the center of the lens while transferring the flakes. Polycarbonate (PC) crystals were dissolved in chloroform (6% by weight) with a magnetic stirrer overnight at 30ºC. A few drops were cast on a glass slide and immediately another glass slide was pressed on it and then slid down to ensure uniform PC film on both glass slides. After 20 minutes of curing, a scotch tape with a window of opening cut out in the middle was used to pick-up the PC film and lay it on the PDMS hemispherical handle, thus completing the process of the "transfer handle" creation. In order to perform the deterministic transfer, the $h$BN on $VO_2$ was put on a temperature controlled hot plate and heated up to 130ºC.

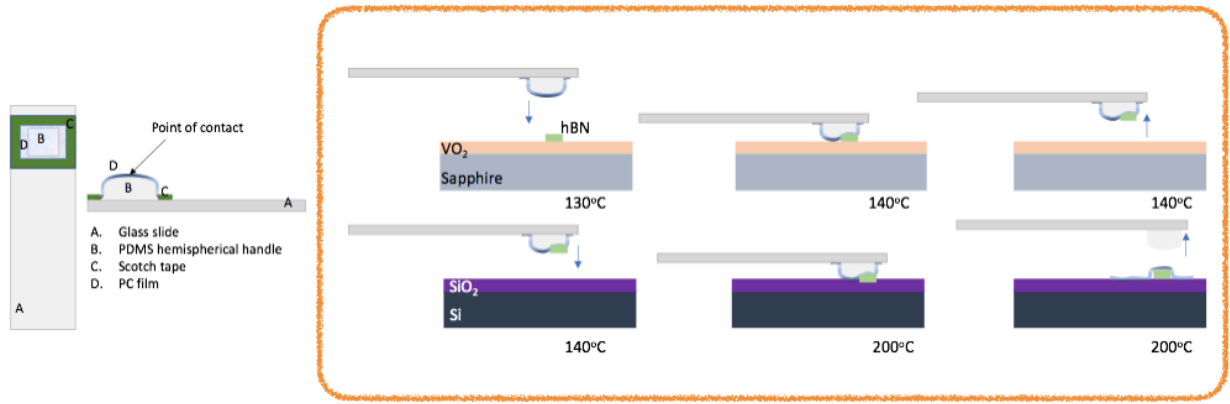

Fig. S13: Transfer Process: Schematic of dry transfer process used to transfer our $h$BN flake on VO$_2$/Sapphire substrate to SiO$_2$/Si substrate.

The PC/PDMS block was slowly brought in contact with the hBN flake of interest using a micro-manipulator and the system was heated slowly up to 140ºC to ensure that the PC film uniformly touches the entire $h$BN flake. At this temperature, the PC is stickier than the VO$_2$ and attracts the $h$BN. Then the system is slowly retracted without lowering the temperature. As the PC film detaches from the VO$_2$ substrate, it lifts the $h$BN flake of interest along with it. It should be noted that while conventionally this is achieved by lowering the temperature rather than the substrate, we found that this reduced the yield in our case because then the VO$_2$ is stickier than the PC at lower temperatures (<130ºC) and the flake does not pick up reliably. Once the $h$BN is on PC/PDMS it can be arbitrarily placed on any substrate with a spatial accuracy of 1um, limited by our transfer setup. In order to "drop" it on to the desired substrate, we bring the hBN flake and substrate in contact at 140ºC. As soon as the PC makes full contact with the substrate, the temperature is taken up to 200ºC and the PC melts and "drops" on the substrate. The PC film can be removed by gentle chloroform and isopropanol treatment sequentially, for 5 minutes. Our transfer process does not involve etching or usage of any harsh chemicals thus preserving the quality of the flakes and also the emitters. Also, our choice of a hemispherical handle allows us greater flexibility in deterministically picking and placing $h$BN flakes, which greatly increases the versatility of this process over other existing methods where the entire substrate is usually etched, and all flakes are transferred to the next substrate. After successful transfer of our flake onto SiO$_2$/Si substrate we performed a confocal PL map of the region containing emitters A and B, to relocate the quantum emitters. We could match spectral and spatial signatures of both emitters (Fig. 5) in the $h$BN flake after transfer onto Si/SiO$_2$. By comparing the PL map before and after transfer it can be observed that the spatial emission patterns before and after transfer are different which can be a result of the dependence of activation of different defect sites in $h$BN on the substrate. This dependence can be attributed to the different optical and electrical properties of the two substrates. The entire transfer process and relocating of the emitters is summarized in Fig. S13.